# The Gender Gap in Science Communication on TikTok and YouTube: How Platform Dynamics Shape the Visibility of Female Science Communicators

**Maider Eizmendi-Iraola[1], Simón Peña-Fernández[2]* and Jordi Morales-i-Gras[3]**

1. University of the Basque Country (UPV/EHU) (ROR: 000xsnr85)
   maider.eizmendi@ehu.eus / https://orcid.org/0000-0002-5894-8238
2. University of the Basque Country (UPV/EHU) (ROR: 000xsnr85)
   simon.pena@ehu.eus / https://orcid.org/0000-0003-2080-3241
3. University of the Basque Country (UPV/EHU) (ROR: 000xsnr85)
   jordi.morales@ehu.eus / https://orcid.org/0000-0003-4173-3609

\* Corresponding author

**Abstract**: Social media platforms facilitate the dissemination of science and access to it. However, gender inequalities in the participation and visibility of communicators persist. This study examined the differences in reach and audience response between YouTube and TikTok from a gender perspective. To do so, the ten most influential science accounts on YouTube and TikTok were selected, with the sample divided equally between men and women, to conduct a comparative study. A total of 4293 videos on TikTok and 4825 on YouTube were analyzed, along with 277,528 comments, considering metrics of views and interaction. The results show that on YouTube, men received more likes and views, while on TikTok, audience response was more balanced. The participation of women on both platforms also had a differential impact, as the number of women engaging with content on YouTube negatively correlated with interaction levels, whereas on TikTok, their impact was slightly positive. In conclusion, TikTok emerges as a more inclusive space for scientific communication, though structural challenges remain on both platforms, encouraging further research into strategies that promote gender equity in online science communication.

## 1. Introduction

The internet has broken down many of the technical and economic barriers that once hindered science communication (Peters et al., 2014), making it easier for the research community to engage in outreach. Whether professionals or casual users, more and more people are turning to these platforms to create and share content. These tools have enabled them to communicate their research quickly and efficiently across the globe, playing a crucial role in the dissemination of scientific information among professionals in the field and also the general public (Coletti et al., 2022). Increasingly, people are turning to the internet as their primary source of scientific information (Dunwoody, 2014), offering an alternative for those interested in science but skeptical of traditional media outlets (Metag, 2020). As a result, science communication has become an essential pathway for acquiring scientific knowledge and literacy. According to the report Trust in science and scientific populism in Spain (FECYT, 2024), 71.3% of respondents reported using the internet to stay informed about science, with online videos surpassing social media as the preferred channel. Notably, the use of social media and video platforms is particularly prominent among young people.

In this context, social media holds immense potential for science communication, thanks to its popularity and the wide range of formats it offers to content creators. Through audiovisual content, the scientific community can showcase, explain, and illustrate information using various techniques and codes (Zeng & Kaye, 2022). Theoretically, social media platforms can also enhance the visibility of female scientists and researchers by helping them overcome the obstacles they have traditionally faced when trying to access mainstream media as expert sources (Mena-Young, 2018; Eizmendi-Iraola & Peña-Fernández, 2023b). However, various studies indicate that women are less engaged as producers of science communication content, both within academic networks (Raffaghelli & Manca, 2023) and on platforms (Welbourne & Grant, 2016), resulting in lower participation and visibility in scientific discourse (Cambronero-Saiz et al., 2021).

This research aims to analyse the interaction between science communicators on social media platforms and the comments they receive, viewed through a gender perspective. To achieve this, the study focuses on leading YouTube and TikTok accounts. YouTube was selected as it is one of the three main platforms for content dissemination and the most commonly used for accessing information about science and research (Metag, 2020). TikTok, on the other hand, has been the most downloaded platform since 2020 (Cheng & Li, 2024). Its younger user base makes it an ideal tool for challenging stereotypes and reshaping perceptions of female scientists through increased visibility.

Based on this approach, this research addressed the following research questions:
RQ1. How do audiences respond to the content produced by leading science communicators on YouTube and TikTok?
RQ2. What gender-based differences can be observed in the interaction received by leading science communicators on YouTube and TikTok?
RQ3. What is the gender distribution of individuals interacting with the content published by prominent science communicators?
RQ4. Is there a correlation between the gender of the content creator and the gender of the individuals engaging with their content?

## 2. Literature Review

The presence of women in science communication through digital platforms has become an increasingly important topic in academic research, primarily due to the growing significance of science communication activities and the gender inequalities that still persist in this field. Regarding the causes of these differences, research suggests that women's participation in science communication is shaped by the barriers they encounter both when entering the scientific field and throughout their career development. Women currently represent 29.3% of the global research workforce (UNESCO, 2018), though this figure varies by country (European Commission, 2019). Beyond these disparities, women in science often face difficulties advancing in their careers, with a higher proportion of women than men leaving academia. As a result, they encounter the so-called "glass ceiling" and suffer from what is known as the "leaky pipeline" phenomenon (Hunt, 2016).

Regarding these obstacles, studies examining the factors behind gender inequality faced by women in the scientific field have highlighted several causes. These include the challenges of balancing academic and research work with caregiving responsibilities (Myers et al., 2020), which are predominantly assigned to women. These circumstances were confirmed by multiple studies conducted during the COVID-19 pandemic, which concluded that the scientific output and outreach activities of female researchers with dependents were significantly affected during this period (Krukowski et al., 2021; Deryugina et al., 2021; Eizmendi-Iraola & Peña-Fernández, 2023a).

However, gender inequalities also stem from structural asymmetries and prevailing androcentric values within both science and academia (Villar-Aguilés & Obiol-Francés, 2022; Read, 2024). In addition, enduring stereotypes persist despite the progress made by women in academic settings (Sugimoto & Larivière, 2023). Generally, men are perceived as leaders and as more analytical, competitive, and independent, while women are more commonly associated with traits such as kindness, warmth, empathy, and helpfulness (Carli et al., 2016). Notably, the traits traditionally linked to men are more closely aligned with how scientific work is typically perceived. In this regard, science communication itself can serve as a pathway for overcoming these obstacles by enhancing visibility and creating role models that promote gender equality.

Studies that have addressed scientific communication emphasize that, in terms of presence, women, in general, tend to be less active on social media platforms (Procter et al., 2010; Cambronero-Saiz et al., 2021). Furthermore, as far as scientific dissemination is concerned, their activity differs in nature, as they are less inclined to use these platforms to showcase personal achievements (Montesi et al., 2019). The topics they address also differ, as the prevalence of women in the dissemination of social sciences content is higher than that of men, who tend to focus more on the basic sciences (Micaletto-Belda et al., 2024). It is also worth noting that the feedback they receive from social media users is unequal; female researchers not only receive fewer responses but also face more hostile comments (Amarasekara & Grant, 2019), and even experience increased attacks after engaging in science communication (FECYT, 2025).

In this context, platforms with a more visual nature like TikTok and YouTube (Steinke et al., 2024b) open up new possibilities in science communication. They make it accessible in terms of content, audience interaction, and global reach, while also enabling innovation in the presentation of materials (Hill et al., 2022). For instance, TikTok, despite not being the ideal format for conveying scientific information with full rigor (Muñoz-Gallego et al., 2024), offers the brevity and novelty that younger audiences demand (De-Casas-Moreno et al., 2024). Thus, the content on TikTok and its viral trends offer an opportunity to increase the visibility of women and other minority groups (Lee & Lee, 2023; Civila & Jaramillo-Dent, 2022). This helps to broaden the public imagination surrounding scientific personnel and to promote scientific vocations among girls and women (Steinke et al., 2024a), as identifying with a role model in the STEM field can lead to changes in stereotypes (Van Camp et al., 2019). At present, the self-conceptual construction of gender is closely linked to the relationship young people have with new information technologies (Popa & Gavriliu, 2015).

## 3. Materials and Methods

To identify the most successful YouTube and TikTok accounts in terms of content production and follower count, the platforms' own search engines were used. By searching the keyword 'Science', accounts focused on this theme were located and ranked according to their number of subscribers. Additionally, three complementary criteria were used to finalize the sample selection, as follows: (1) the accounts had to focus exclusively on science-related content; (2) the accounts had to be personal rather than collective or organization-run; and (3) the videos had to feature the science communicator appearing in person. These criteria were applied to ensure the consistency and relevance of the sample. Focusing exclusively on science-related content guaranteed thematic coherence across accounts. Selecting individual creators, rather than organizations or collectives, allowed the analysis of personal communication styles and gender representation, which are central to this study. Finally, including only videos in which the science communicator appears on screen ensured that this analysis reflects not only content but also visual presence and performance, both of which are key to understanding audience participation, visibility, and the construction of identity in digital science communication.

Table 1. Science communication accounts.

| | TikTok | | | YouTube | |
|---|---|---|---|---|---|
| Channel | Gender | No. of Followers | Channel | Gender | No. of Followers |
| anacnd | female | 2,900,000 | AsapSCIENCE | male | 10,700,000 |
| arsen | male | 4,500,000 | Crazy Medusa | female | 224,000 |
| asapscience | male | 1,200,000 | minutephysics | male | 5,840,000 |
| ladyscience | female | 870,200 | Physics Girl | female | 3,360,000 |
| melscience | male | 83,000 | Primrose Kitten | female | 250,000 |
| realsciencebob | male | 680,700 | Sally Le Page | female | 85,300 |
| science.bae | female | 140,400 | Science with Hazel | female | 94,100 |
| science.is.magik | female | 522,900 | SmarterEveryDay | male | 11,600,000 |
| scienceirl | male | 15,000 | Veritasium | male | 17,400,000 |
| sciencewithana | female | 3,500,000 | Vsauce | male | 23,600,000 |

Following these criteria, a total of 20 accounts were selected, 10 from each of the analysed platforms. In each of them, the five most-followed male-led accounts and the five most-followed female-led accounts were chosen (Table 1). The decision to include only the top ten accounts per platform was based not only on analytical manageability but also on the observed distribution of follower counts. From the 11th position onwards, there was a noticeable drop in the number of followers, which could have affected the representativeness and visibility of the accounts in the context of science communication. Therefore, focusing on the top 10 ensured that the sample consisted of the most influential and widely followed science communicators, allowing more robust comparison across platforms.

TikTok data were downloaded using the Ensembledata API, while YouTube data were retrieved via the YouTube Data Tools API from the University of Amsterdam. The analysis covered a 12-month period throughout the year 2022. To approach an analysis of audience response, several of its key indicators—namely views, likes, and comments—were examined across 4293 TikTok videos and 4825 YouTube videos produced by the selected accounts. The number of views and likes was recorded for each of the videos, while comments were analyzed through gender attribution using the gender-guesser Python library, applied to the usernames of commenters (https://github.com/lead-ratings/gender-guesser, accessed on 16 March 2024). The gender-guesser script classifies names into six possible categories: "male," "mostly male," "female," "mostly female," "androgynous," or "unknown." For the present analysis, comments classified as "male" or "mostly male" were grouped into the "male" category, and similarly, comments classified as "female" or "mostly female" were grouped into the "female" category. Comments classified as "androgynous" or "unknown" were discarded. Only clearly attributable gender classifications (36.1% of the total) were used, while the remaining cases were discarded.

In total, 256,488 comments on YouTube and 21,040 comments on TikTok were analyzed. It is important to note that the analysis did not control for multiple comments by the same user, as the focus was on describing overall audience behavior rather than individual user activity.

## 4. Results

*4.1. Follower Base on YouTube, Virality on Tiktok*

Among the leading science communicators, YouTube has the broadest reach. Overall, the analyzed videos had an average of 1,513,260 views, compared to 364,228 on TikTok. This difference is likely influenced by YouTube's longer trajectory (YouTube was launched in 2005, while TikTok was introduced in 2017), which allows users to build a more robust subscriber base (7.3 million on average, compared to 1.4 million on TikTok among the accounts that make up the sample) (Table 1). Both platforms have a similar video view-to-subscriber ratio (Table 2 and Table 3). The videos on YouTube achieved an average of 0.21 views per subscriber, while on TikTok, this figure increased slightly to 0.25. This gave YouTube an initial advantage due to its larger follower base, which translated into a greater overall number of views and likes per video.

TikTok, by contrast, stands out for the virality of its content. In the analysis conducted, leading science communicators on the platform receive, on average, one like for every 7.5 views. On YouTube, however, this figure rises to 45.1 views per like. In other words, TikTok content requires far fewer viewers to generate a like, suggesting a higher level of immediate engagement and a more reactive audience compared to YouTube. However, it should be noted that differences in user engagement may also be influenced by platform-specific interfaces and affordances, as well as distinct audience cultures surrounding YouTube and TikTok. This recommendation system is particularly relevant as it facilitates content discovery beyond a user's immediate network.

This dual platform trend is well illustrated by the case of the only channel in the study present on both platforms: AsapSCIENCE. On YouTube, where it has a substantial subscriber base of 10.7 million, its videos achieved an average of 4,179,070 views and 75,138 likes. This translates to 0.39 views per subscriber and 56.2 views per like. In contrast, on TikTok, where its follower base is nearly one-tenth of its YouTube audience (1.2 million), its videos generated 0.24 views per subscriber but achieved one like for every 5.7 views. This comparison highlights YouTube's advantage in sheer reach due to its larger subscriber base, while TikTok's algorithm-driven engagement fosters a significantly higher like-to-view ratio, reinforcing its reputation as a platform with stronger content virality.

Table 2. Views and likes of TikTok science communication accounts.

| Channel | TikToker's Gender | Videos | Sum of Views | Average Views | Sum of Likes | Average Likes |
|---|---|---|---|---|---|---|
| anacnd | female | 253 | 396,156,296 | 1,565,835 | 44,900,000 | 177,470 |
| arsen | male | 133 | 246,144,400 | 1,850,710 | 48,800,000 | 366,917 |
| asapscience | male | 752 | 216,676,025 | 288,133 | 37,700,000 | 50,133 |
| sciencewithana | female | 444 | 207,374,187 | 467,059 | 31,300,000 | 70,496 |
| ladyscience | female | 1162 | 168,128,507 | 144,689 | 7,700,000 | 6627 |
| melscience | male | 492 | 148,848,588 | 302,538 | 17,600,000 | 35,772 |
| science.is.magik | female | 99 | 58,692,499 | 592,854 | 6,300,000 | 63,636 |
| science.bae | female | 452 | 54,942,962 | 121,595 | 3,500,000 | 7743 |
| scienceirl | male | 472 | 38,897,022 | 82,409 | 6,200,000 | 13,135 |
| realsciencebob | male | 34 | 27,768,699 | 816,726 | 4,600,000 | 135,294 |

Table 3. Views and likes of YouTube science communication accounts.

| Channel | YouTubers' Gender | Videos | Sum of Views | Sum of Likes | Average Views | Average Likes |
|---|---|---|---|---|---|---|
| Vsauce | male | 389 | 2,186,888,939 | 48,897,526 | 5,975,106 | 134,190 |
| AsapSCIENCE | male | 412 | 1,713,418,773 | 30,505,920 | 4,179,070 | 75,138 |
| Veritasium | male | 331 | 1,640,277,664 | 40,468,652 | 4,970,538 | 122,632 |
| SmarterEveryDay | male | 353 | 1,038,687,823 | 21,519,437 | 2,950,818 | 61,135 |
| minutephysics | male | 260 | 494,077,778 | 9,692,608 | 1,907,636 | 37,279 |
| Physics Girl | female | 193 | 185,436,246 | 9,903,857 | 960,809 | 25,409 |
| Primrose Kitten | female | 2036 | 32,177,219 | 627,966 | 15,812 | 308 |
| Science w. Hazel | female | 635 | 5,226,112 | 122,558 | 8230 | 193 |
| Crazy Medusa | female | 129 | 3,031,658 | 117,489 | 23,501 | 911 |
| Sally Le Page | female | 87 | 2,258,678 | 93,973 | 25,962 | 1080 |

*4.2. Male Skew on YouTube, Gender Balance on TikTok*

From a gender perspective, the platform distribution reveals a noticeable imbalance among the most prominent science communicators. On YouTube, the five male-led accounts dominate, amassing 69.1 million subscribers and 7.1 billion views. In contrast, the female-led accounts have accumulated only 4 million subscribers and 228 million views. On TikTok, the trend is reversed; women have 7.9 million followers and 885 million views, compared to 6.5 million followers and 678 million views for men.

Although the sample size is limited and does not allow broad generalizations about the overall presence of men and women in science communication on these platforms, the disparity is stark in the analyzed sample of the ten leading accounts by gender. YouTube remains highly male-dominated, whereas TikTok presents a more balanced gender distribution when considering science communication creators. The dominance of men among YouTube's most popular science communicators is overwhelming, as they account for 96.9% of total views, 96% of comments, and 93.3% of likes. In contrast, TikTok's numbers are far more balanced, with women generating 56.6% of total views and 44.9% of likes among the leading science communication accounts.

Interestingly, audience response indicators follow an inverse trend. On YouTube, women achieve a stronger audience response, with one like every 21 views and one comment every 421 views. For men, these figures increase to 46.8 views per like and 538.2 views per comment, indicating a lower interaction rate per view. On TikTok, however, the pattern flips; male-led accounts require 5.9 views to earn a like, while female-led accounts need 9.4 views per like.

These findings suggest that while male science communicators dominate YouTube in terms of reach and total audience response, female communicators tend to foster a higher level of interaction per view on YouTube. Meanwhile, TikTok offers a more level playing field in terms of visibility, though men achieve stronger participation metrics. This comparison highlights YouTube's advantage in sheer reach due to its larger subscriber base, while TikTok's algorithm-driven engagement fosters a significantly higher like-to-view ratio, reinforcing its reputation as a platform with stronger content virality.

*4.3. The Gender of Comments*

Given the general interaction data, it is worth examining whether there is a differential attitude between men and women when engaging with the content of YouTubers and TikTokers. Based on the users whose gender could be identified using an algorithm (36.1% of the total), the previously observed user distribution differences across both platforms extend to the commenters. On TikTok, women make up 41.6% of commenters on female TikTokers' content and 39.6% of commenters on male TikTokers' content. While not perfectly gender-balanced, these figures indicate a relatively even distribution of audience response across both cases (Figure 1). In contrast, the gender gap widens significantly on YouTube. Women account for 36.2% of comments on female YouTubers' videos, but their presence drops to just 16.3% on male YouTubers' videos, which also tend to have the highest view counts (Figure 2). Overall, on this platform, the greater the number of views and total comments a video receives, the lower the proportion of comments from women.

The analysis of the available data on TikTok (Figure 3) indicated that the observed and expected values are quite similar, suggesting that the creator's gender does not drastically influence the profile of commenters. However, the chi-square test ($\chi^2 = 6.08$, $p = 0.014$) confirmed that the association between the two variables was statistically significant, though its magnitude was extremely low, as evidenced by the Phi coefficient (0.0170) and Cramér's V (0.0170).

Figure 1. Comments per channel on TikTok by imputed gender of commenters.

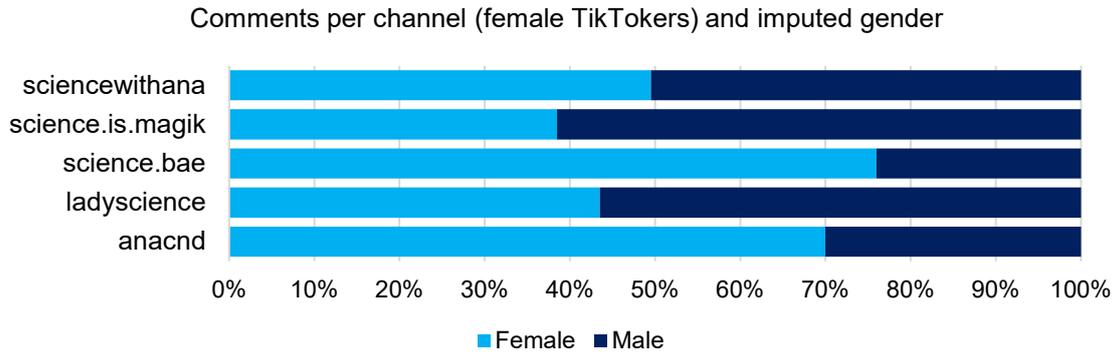

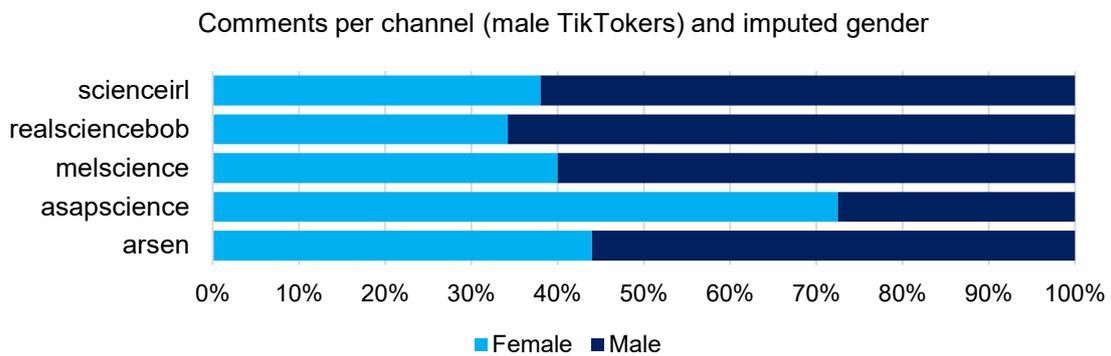

Figure 2. Comments per channel on YouTube by imputed gender of commenters.

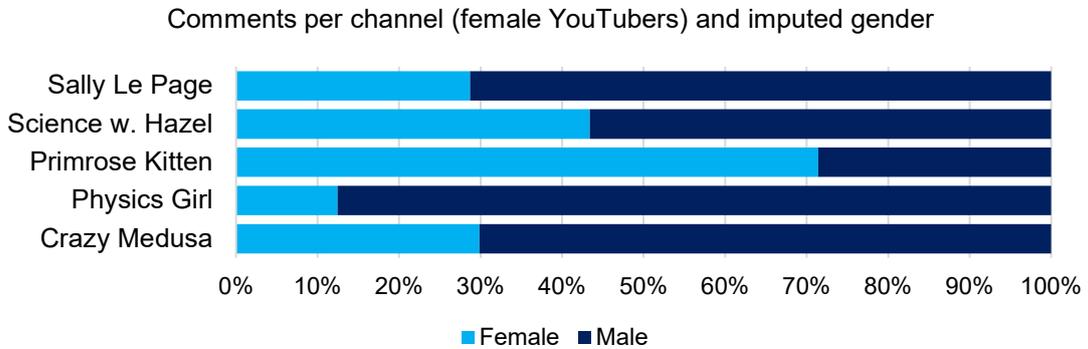

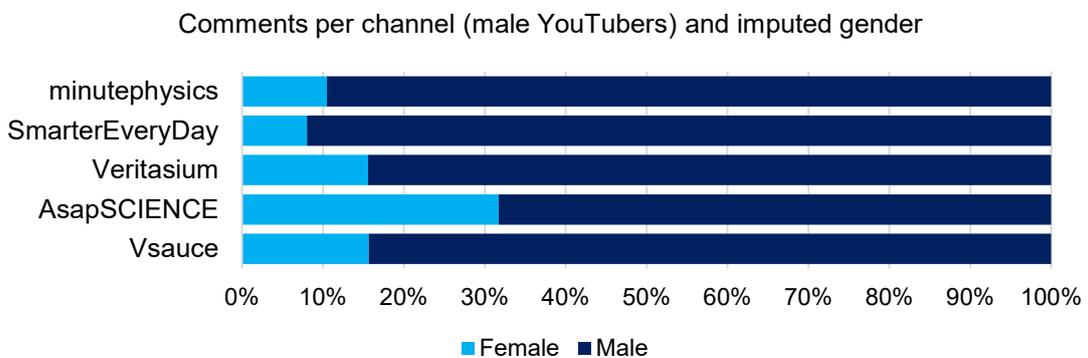

Figure 3. Relationship between TikTok creator's gender and commenters' gender.

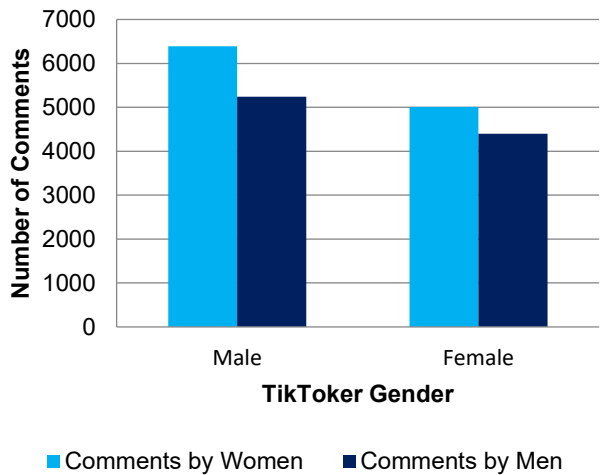

**Contingency Tables**

| TikToker Gender | | Commenter Gender | | |
|---|---|---|---|---|
| | | Female | Male | Total |
| Female | Observed | 5006 | 4400 | 9406 |
| | Expected | 5095 | 4311 | 9406 |
| Male | Observed | 6390 | 5244 | 11,634 |
| | Expected | 6301 | 5333 | 11,634 |
| Total | Observed | 11,396 | 9644 | 21,040 |
| | Expected | 11,396 | 9644 | 21,040 |

**χ² Tests**

| | Value | df | p |
|---|---|---|---|
| χ² | 6.08 | 1 | 0.014 |
| N | 21040 | | |

**Nominal**

| | Value |
|---|---|
| Phi-coefficient | 0.0170 |
| Cramer's V | 0.0170 |

Overall, comments were relatively evenly distributed between men and women, with no strong preference for interacting with creators of the same gender. While slight deviations from expected values existed—men tended to comment slightly more on videos by other men, and women on videos by other women—these differences were minimal and practically insignificant. The statistical significance suggests that some pattern may exist in comment distribution, but the effect is so weak that it is unlikely to have any substantial behavioral implications on the platform. Specifically, although the majority of comments were made by men, in three of the ten analyzed cases (two female TikTokers and one male), there were more comments from women than from men. Therefore, in the case of leading science communicators on TikTok, while the data show a slight tendency toward same-gender interaction, the difference is so small that it can be considered negligible in practical terms. This suggests that, in terms of comment participation, TikTok provides an environment where the creator's gender is not a determining factor in audience interaction dynamics.

In the case of YouTube (Figure 4), the observed and expected values showed more noticeable differences than on TikTok, suggesting that the creator's gender may have had a clearer influence on the composition of the commenting audience. The chi-square test ($\chi 2 = 35.1$, $p < 0.001$) indicated that the relationship between the two variables was highly statistically significant. This means that the distribution of comments was not entirely random in relation to the creator's gender. However, despite this statistical significance, the Phi coefficient (0.0117) and Cramér's V (0.0117) remained extremely low, indicating that the strength of the association was very weak.

Figure 4. Relationship between YouTube creator's gender and commenters' gender.

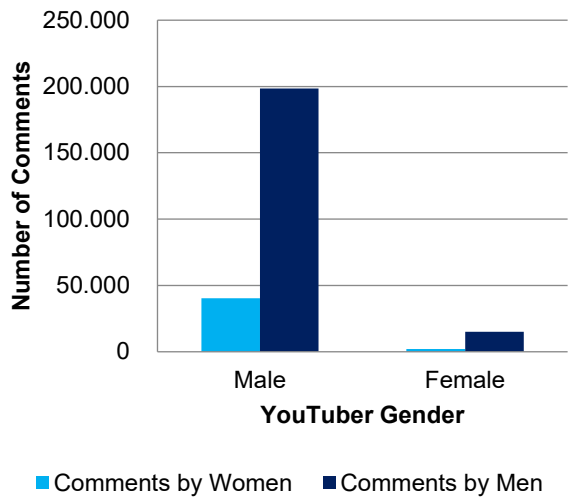

| Contingency Tables | | | | |
|---|---|---|---|---|
| | | Commenter Gender | | |
| Youtuber Gender | | Female | Male | Total |
| Female | Observed | 2691 | 15,079 | 17,770 |
| | Expected | 2975 | 14,795 | 17,770 |
| Male | Observed | 40,255 | 198,463 | 238,718 |
| | Expected | 39,971 | 198,747 | 238,718 |
| Total | Observed | 42,946 | 213,542 | 256,488 |

| $\chi^2$ Tests | | | |
|---|---|---|---|
| | Value | df | p |
| $\chi^2$ | 35.1 | 1 | < 0.001 |
| N | 256488 | | |

| Nominal | |
|---|---|
| | Value |
| Phi-coefficient | 0.0117 |
| Cramer's V | 0.0117 |

In terms of user behavior, the data suggest that male creators receive a higher proportion of comments from men, whereas female creators receive a more balanced distribution of comments, albeit with a significantly lower total number of interactions. However, the magnitude of these differences remained so small that, from a practical standpoint, the impact of this relationship is almost negligible. While the analysis revealed a statistically significant association between the creator's gender and the commenter's gender, the actual strength of this relationship is minimal. In practical terms, this suggests that while there may be a slight tendency for men to comment more on videos by other men, the creator's gender is not a decisive factor in shaping interaction dynamics within YouTube comments. On this platform, the presence of women in the comments is higher when the science communicator is a woman, provided it is a small-scale project with limited reach. In larger projects, this effect fades, with the presence of women even exceeding expectations in videos featuring male creators.

In other words, on both platforms, while there is a statistically significant relationship between the gender of science communicators and the gender of their commenters, its practical impact is limited. The observed difference was too small to be considered a key factor in shaping interaction dynamics. Therefore, although gender may slightly influence comment distribution, other contextual factors are likely to play a more substantial role in audience participation.

*4.4. Gender Dynamics Interaction*

The analysis of the correlation values for the proportion of women commenting in relation to interaction metrics on TikTok, specifically assessing whether the number of female commenters influenced the interaction levels, revealed a moderate and statistically significant association between the number of female commenters and the number of comments received (r = 0.313, p < 0.05). This suggests that videos with greater female participation in the comments tend to generate more discussion, although the effect is moderate and not necessarily determinative of audience engagement (Figure 5). These data demonstrate that, despite a slight trend towards a higher proportion of female commenters, virality and audience response on TikTok are primarily driven by other factors, such as video content, audience preferences, and the platform's algorithm, rather than the proportion of women participating in the conversation, although this is a variable that should be taken into account.

Figure 5. Correlation matrix between the proportion of female commenters and interaction metrics on TikTok and YouTube.

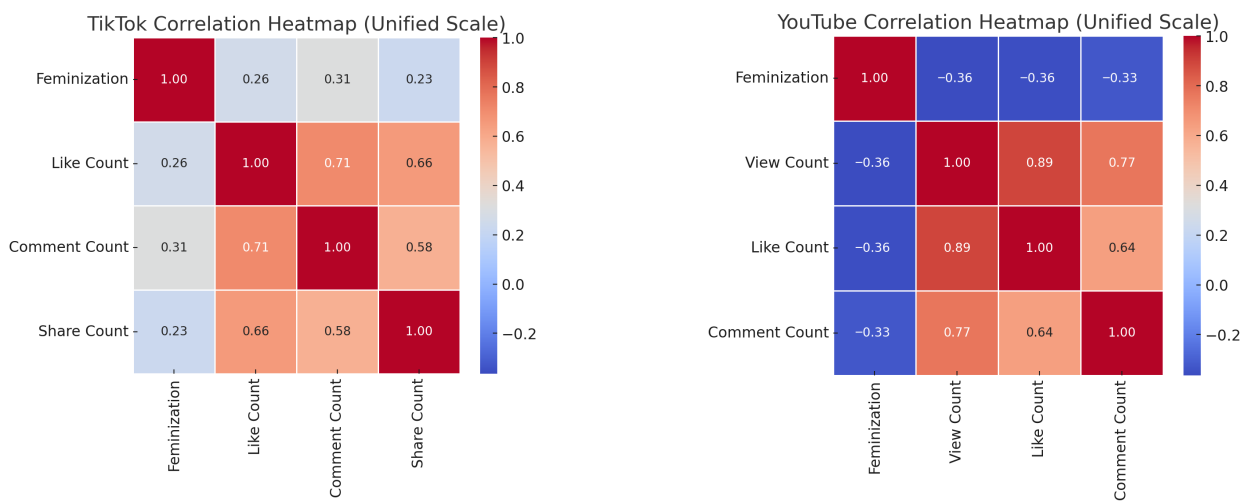

In contrast, the data regarding the presence of women among commenters on YouTube show a different trend. On this platform, the analysis of interactions in relation to the proportion of female commenters revealed negative and statistically significant correlations. Specifically, videos with a higher proportion of women among commenters tended to receive fewer views (r = −0.364, p < 0.05), fewer likes (r = −0.356, p < 0.05), and fewer comments (r = −0.332, p < 0.05). Although these correlations were moderate in strength, they were statistically significant and consistent in direction, suggesting that greater female engagement in the comment section is associated with reduced audience interaction on this platform.

Taken together, these findings indicate distinct patterns across platforms. On TikTok, a higher proportion of female commenters is associated with increased comment activity, whereas on YouTube, it corresponds consistently with lower levels of views, likes, and comments. These results highlight clear differences in audience interaction dynamics between the two platforms.

## 5. Discussion

Social media and platforms like YouTube and TikTok are key resources for increasing the reach of science communication and fostering dialogue with audiences (Oliver et al., 2023). They provide an agile way to transfer knowledge in engaging and easily consumable formats (Hill et al., 2022). From a gender perspective, these platforms also offer women a new opportunity to gain visibility, reducing the gap in the social perception of women in science and challenging stereotypes (Huber & Quesada-Baena, 2023).

YouTube and TikTok offer differentiated opportunities for science communication. While YouTube offers a large follower base to reach wide audiences through a robust subscriber system, TikTok has a low entry barrier and an algorithm that allows content to reach more diverse audiences (Zawacki et al., 2022). The different sociodemographic characteristics of their users, with a greater number of young people on TikTok, also influence how the content is consumed (Pew Research Center, 2024).

In the case of science communication, we encounter a clear male dominance on YouTube and greater gender parity on TikTok. In particular, while male creators dominate reach and view metrics on YouTube, TikTok emerges as a more accessible platform for women, allowing for greater visibility and a more equitable distribution of the audience. This may be related to the fact that the number of female users on TikTok is higher than on other platforms, although this trend may vary across countries (IAB, 2024; Gutiérrez & Ramírez, 2022). Furthermore, the data reveal that interaction levels vary based on the creator's and user's gender. On YouTube, the female audience is significantly smaller on male-led channels, whereas on TikTok, there is no strong association between the creator's gender and the composition of the comments.

Regarding TikTok, the analysis revealed a moderate and statistically significant association between the proportion of female commenters and the number of comments received. In any case, this may be in line with other studies suggesting that women are more likely to engage in online discussions, for instance, in relation to political issues (Lybeck et al., 2024). In contrast, on YouTube, a higher proportion of women among commenters showed a negative correlation with the numbers of views, likes, and comments, suggesting that content attracting a higher proportion of female commenters may have a smaller reach on the platform. The greater the presence of women in the conversation on YouTube, the lower the viewership and virality of the videos. This may be related to the fact that, traditionally, male scientists have been attributed greater credibility than their female counterparts. However, recent studies have revealed the role played by perceived likability, which audiences more commonly associate with women, in shaping perceptions of competence (Hubner & Bullock, 2024).

These findings reinforce the idea that scientific communication on social media is influenced not only by algorithmic and format-related factors (Martín Neira et al., 2023; Velarde-Camaqui et al., 2024), but also by gender dynamics, which affect the participation and impact of communicators in a differentiated way on both platforms. The results confirm previous findings on the gender gap in science communication on social media (Amarasekara & Grant, 2019; Cambronero-Saiz et al., 2021) while also providing new insights into how this gap varies depending on the platform used. While YouTube continues to reflect a traditionally male-dominated structure, TikTok emerges as a platform with less gender bias, suggesting that new digital formats could play a role in balancing the presence of women in science communication (Huber & Quesada-Baena, 2023).

This study presents certain limitations, particularly concerning the sample size and the focus on two platforms and creators who have already achieved a significant following. For its part, the gender-guesser Python library assigns gender based on first names using a database primarily derived from Anglo-Saxon sources. It performs well with common names in Western contexts but has limitations with non-Anglo names or pseudonyms, which are common on social media. It does not identify gender identity, but rather the gender historically associated with a name.

Future research could investigate the perceptions of audiences who consume this content and assess the effects of such narratives on their attitudes. In addition, interviews and focus groups with content creators on TikTok and YouTube could offer valuable insights into their experiences, motivations, and goals.

**Funding**: This work is part of the scientific output of the Consolidated Research Group of the Basque University System, Gureiker (IT1496-22).

## References

Amarasekara, I., & Grant, W. (2019). Exploring the YouTube science communication gender gap: A sentiment analysis. Public Understanding of Science, 28(1), 68–84.

Cambronero-Saiz, B., Segarra-Saavedra, J., & Cristófol-Rodríguez, C. (2021). Análisis desde la perspectiva de género del engagement de los principales YouTubers de divulgación científica. Cuestiones de Género: De la Igualdad y la Diferencia, 16, 511–525.

Carli, L., Alawa, L., Lee, Y., Zhao, B., & Kim, E. (2016). Stereotypes about gender and science: Women≠scientists. Psychology of Women Quarterly, 40(2), 244–260.

Cheng, Z., & Li, Y. (2024). Like, comment, and share on TikTok: Exploring the effect of sentiment and second-person view on the user engagement with TikTok news videos. Social Science Computer Review, 42(1), 201–223.

Civila, S., & Jaramillo-Dent, D. (2022). #Mixedcouples on TikTok: Performative hybridization and identity in the face of discrimination. Social Media + Society, 8(3), 20563051221122464.

Coletti, A., McGloin, R., Oeldorf-Hirsch, A., & Hamlin, E. (2022). Science communication on social media. The Journal of Social Media in Society, 11(2), 236–263.

De-Casas-Moreno, P., Blanco-Sánchez, T., & Parejo-Cuéllar, M. (2024). El efecto TikTok como nuevo entorno para la divulgación científica. Observatorio (OBS)*, 18(3), 36–53.

Deryugina, T., Shurchkov, O., & Stearns, J. (2021). COVID-19 disruptions disproportionately affect female academics. AEA Papers and Proceedings, 111, 164–168.

Dunwoody, S. (2014). Science journalism: Prospects in a digital age. In M. Bucchi, & B. Trench (Eds.), Routledge handbook of public communication of science and technology (pp. 27–39). Routledge.

Eizmendi-Iraola, M., & Peña-Fernández, S. (2023a). Gender stereotypes make women invisible: The presence of female scientists in the media. Social Sciences, 12(1), 30.

Eizmendi-Iraola, M., & Peña-Fernández, S. (2023b). ¿Tiene género la divulgación científica? Análisis de los artículos publicados en The Conversation sobre el COVID-19. TechnoReview, 13(1), 1–12.

European Commission, Directorate-General for Research and Innovation. (2019). She figures 2018. Publications Office of the European Union.

Fundación Española para la Ciencia y la Tecnología (FECYT). (2024). Confianza en la ciencia y populismo científico en España. Ministerio de Ciencia, Innovación y Universidades.

Fundación Española para la Ciencia y la Tecnología (FECYT). (2025). Participación de las científicas como fuentes expertas en los medios: Motivaciones y obstáculos. Fundación Española para la Ciencia y la Tecnología (FECYT).

Gutiérrez, P., & Ramírez, A. (2022). El deseo de los menores por ser youtuber y/o influencer. Narcisismo como factor de influencia. Pixel-Bit, 63, 227–255.

Hill, V. M., Grant, W. J., McMahon, M. L., & Singhal, I. (2022). How prominent science communicators on YouTube understand the impact of their work. Frontiers in Communication, 7, 1014477.

Huber, B., & Quesada-Baena, L. (2023). Women scientists on TikTok: New opportunities to become visible and challenge gender stereotypes. Media and Communication, 11(1), 240–251.

Hubner, A. Y., & Bullock, O. M. (2024). Why science should have a female face: Female experts increase liking, competence, and trust in science. Science Communication, 47(4), 527–552.

Hunt, J. (2016). Why do women leave science and engineering? ILR Review, 69(1), 199–226.

IAB Spain. (2024). Estudio de redes sociales 2024. Available online: https://goo.su/w4BUk2 (accessed on 16 March 2024).

Krukowski, R., Jagsi, R., & Cardel, M. (2021). Academic productivity differences by gender and child age in science, technology, engineering, mathematics, and medicine faculty during the COVID-19 pandemic. Journal of Women's Health, 30(3), 341–347.

Lee, J. J., & Lee, J. (2023). #StopAsianHate en TikTok: Mujeres asiáticas/estadounidenses crean espacios para liderar contranarrativas y formar una comunidad asiática ad hoc. Redes Sociales + Sociedad, 9(1), 205630512311575.

Lybeck, R., Koiranen, I., & Koivula, A. (2024). From digital divide to digital capital: The role of education and digital skills in social media participation. Universal Access in the Information Society, 23, 1657–1669.

Martín Neira, J. I., Trillo-Domínguez, M., & Olvera-Lobo, M. D. (2023). De la televisión a TikTok: Nuevos formatos audiovisuales para comunicar ciencia. Comunicación y Sociedad, 20, e8441.

Mena-Young, M. (2018). Mujeres científicas en la prensa: Análisis de reportajes de ciencia en diarios de España, México y Costa Rica. Posgrado y Sociedad, 16(1), 2–15.

Metag, J. (2020). What drives science media use? Predictors of media use for information about science and research in digital information environments. Public Understanding of Science, 29(6), 561–578.

Micaletto-Belda, J. P., Morejón-Llamas, N., & Martín-Ramallal, P. (2024). TikTok como plataforma educativa: Análisis de las percepciones de los usuarios sobre los contenidos científicos. Revista Mediterránea de Comunicación, 15(1), 97–144.

Montesi, M., Villaseñor Rodríguez, I., & Bittencourt dos Santos, F. (2019). Presencia, actividad, visibilidad e interdisciplinariedad del profesorado universitario de Documentación en los medios sociales: Una perspectiva de género. Revista Española de Documentación Científica, 42(4), e246.

Muñoz-Gallego, A., Giri, L., Nahabedian, J. J., & Rodríguez, M. (2024). Narrativas audiovisuales en Tik Tok: Nuevos desafíos para la comunicación pública de la ciencia y la tecnología. Revista Mediterránea de Comunicación/Mediterranean Journal of Communication, 15(1), 145–161.

Myers, K., Tham, W., Yin, Y., Cohodes, N., Thursby, J., Thursby, M., Schiffer, P., Walsh, J., Lakhani, K., & Wang, D. (2020). Unequal effects of the COVID-19 pandemic on scientists. Nature Human Behaviour, 4(9), 880–883.

Oliver, E., Redondo-Sama, G., de Aguileta, A. L., & Burgues-Freitas, A. (2023). Research agenda to engage citizens in science through social media communicative observations. Humanities and Social Sciences Communications, 10, 447.

Peters, P., Dunwoody, S., Allgaier, J., Lo, Y. Y., & Brossard, D. (2014). Public communication of science 2.0: Is the communication of science via the "new media" online a genuine transformation or old wine in new bottles? EMBO Reports, 15(7), 749–753.

Pew Research Center. (2024). Social media fact sheet. Available online: https://www.pewresearch.org/internet/fact-sheet/social-media/ (accessed on 16 March 2024).

Popa, D., & Gavriliu, D. (2015). Gender representations and digital media. Procedia-Social and Behavioral Sciences, 180, 1199–1206.

Procter, R., Williams, R., Stewart, J., Poschen, M., Snee, H., Voss, A., & Asgari-Targhi, M. (2010). Adoption and use of Web 2.0 in scholarly communications. Philosophical Transactions of the Royal Society A: Mathematical, Physical, and Engineering Sciences, 368(1926), 4039–4056.

Raffaghelli, J. E., & Manca, S. (2023). Exploring the social activity of open research data on ResearchGate: Implications for the data literacy of researchers. Online Information Review, 47(1), 197–217.

Read, B. (2024). Gender equity in academic knowledge production: The influence of politics, power and precarity. European Educational Research Journal, 24(3), 394–407.

Steinke, J., Gilbert, C., Coletti, A., Levin, S. H., Suk, J., & Oeldorf-Hirsch, A. (2024a). Mujeres en STEM en TikTok: Impulsando la visibilidad y la voz a través de la expresión de la identidad STEM. Redes Sociales + Sociedad, 10(3).

Steinke, J., Gilbert, C., Opat, K., & Landrum, A. R. (2024b). Fostering inclusive science media: Insights from examining the relationship between women's identities and their anticipated engagement with Deep Look YouTube science videos. PLoS ONE, 19(8), e0308558.

Sugimoto, C., & Larivière, V. (2023). Equity for women in science: Dismantling systemic barriers to advancement. Harvard University Press.

UNESCO. (2018). Informe de la UNESCO sobre la Ciencia, hacia 2030: Informe regional de América Latina y el Caribe. UNESCO. Available online: www.unesco.org (accessed on 16 March 2024).

Van Camp, A. R., Gilbert, P. N., & O'Brien, L. T. (2019). Testing the efects of a role model intervention on women's STEM outcomes. Social Psychology of Education, 22(3), 649–671.

Velarde-Camaqui, D., Viehmann, C., Díaz, R., & Valerio-Ureña, G. (2024). Características de los videos que favorecen el engagement de los divulgadores científicos en TikTok. Revista Latina de Comunicación Social, 82, 1–18.

Villar-Aguilés, A., & Obiol-Francés, S. (2022). Academic career, gender and neoliberal university in Spain: The silent precariousness between publishing and care-giving. British Journal of Sociology of Education, 43(4), 623–638.

Welbourne, D. J., & Grant, W. J. (2016). Science communication on YouTube: Factors that affect channel and video popularity. Public Understanding of Science (Bristol, England), 25(6), 706–718.

Zawacki, E. E., Bohon, W., Johnson, S., & Charlevoix, D. J. (2022). Exploring TikTok as a promising platform for geoscience communication. Geoscience Communication, 5, 363–380.

Zeng, J., & Kaye, D. B. V. (2022). From content moderation to visibility moderation: A case study of platform governance on TikTok. Policy & Internet, 14, 79–95.